\documentclass{llncs}

\usepackage{natbib}
\usepackage{graphicx}

\usepackage{fancyhdr}
\usepackage{lastpage}
 
\begin{document}

\title{A short review and primer on event-related potentials in human computer interaction applications}
\author{Minna Huotilainen\inst{1} \and Benjamin Cowley\inst{2,3} \and Lauri Ahonen\inst{2}}
\institute{Finnish Institute of Occupational Health, Helsinki, Finland
\and
Quantitative Employee unit, Finnish Institute of Occupational Health,\\
\email{benjamin.cowley@ttl.fi},\\
POBox 40, Helsinki, 00250, Finland
\and
Cognitive Brain Research Unit, Institute of Behavioural Sciences, University of Helsinki, Helsinki, Finland}

\maketitle              

\begin{abstract}
The application of psychophysiology in human-computer interaction is a growing field with significant potential for future smart personalised systems. Working in this emerging field requires comprehension of an array of physiological signals and analysis techniques. 

Event-related potentials, termed ERPs, are a stimulus- or action-locked waveform indicating a characteristic neural response. ERPs derived from electroencephalography have been extensively studied in basic research, and have been applied especially in the field of brain-computer interfaces. For ecologically-valid settings there are considerable challenges to application, however recent work shows some promise for ERPs outside the lab. Here we present a short review on the application of ERPs in human-computer interaction. 

This paper aims to serve as a primer for the novice, enabling rapid familiarisation with the latest core concepts. We put special emphasis on everyday human-computer interface applications to distinguish from the more common clinical or sports uses of psychophysiology.

This paper is an extract from a comprehensive review of the entire field of ambulatory psychophysiology, including 12 similar chapters, plus application guidelines and systematic review. Thus any citation should be made using the following reference:

{\parshape 1 2cm \dimexpr\linewidth-1cm\relax
B. Cowley, M. Filetti, K. Lukander, J. Torniainen, A. Henelius, L. Ahonen, O. Barral, I. Kosunen, T. Valtonen, M. Huotilainen, N. Ravaja, G. Jacucci. \textit{The Psychophysiology Primer: a guide to methods and a broad review with a focus on human-computer interaction}. Foundations and Trends in Human-Computer Interaction, vol. 9, no. 3-4, pp. 150--307, 2016.
\par}

\keywords{event related potentials, electroencephalography, psychophysiology, human-computer interaction, primer, review}

\end{abstract}

\section{Introduction}

Event-related brain potentials are well-known and widely studied signatures of brain activity that are time-locked to a specific stimulus event, such as the beginning of a sound or the presentation of an image. Distinct from EEG oscillations, ERPs are direct representations of the time domain. These potentials are extracted from the raw EEG signal by averaging over tens to hundreds of EEG time periods (epochs) of fixed duration and are offset with respect to the time stamp for the event \citep{Luck2014}. Some ERP components are so strong that with a large enough number of channels and careful investigation of the spatial and spectral characteristics of the ERP, they can be recognised even without averaging \citep{Delorme2004}. Such `single-trial' ERPs, together with ERPs evoked by more naturalistic stimuli (for example, continuous ecologically valid sounds), constitute the main focus of this section. The broader subject of ERPs is quite vast; for further reading on the topic, see \cite{Kappenman2011}.

Some ERPs, especially those with low latencies, typically remain unaffected by cognitive workload, mental effort, vigilance, or affective processes, while others, especially those related to higher cognitive functions and with higher latencies, can be used reliably as indicators of mental state and task difficulty in HCI settings \citep{Nittono2003}.

\section{Background}

In electrophysiology, cognitive processes are traditionally studied via examination of high-latency neural responses of the ERP that are extracted from the continuous EEG by signal averaging. One can refer to ERPs in terms of their polarity (positive or negative) and the latency after the trigger event. Customarily in plots of ERPs, positive values are positioned below the \emph{x}-axis, and latencies are denoted canonically, not exactly. Hence, for example, two common ERPs are designated as N1 and P3 (these are also referred to as the N100 and P300).

Early-going or `fast' ERPs tend to be related to orienting responses, so the N1 ERP is a fast neuronal response to stimuli. The ERPs associated with cognition typically peak hundreds of milliseconds after the onset of an event and originate in connected cortical areas \citep{savers_mechanism_1974}. Increases in task difficulty, a lack of cognitive resources, and higher cognitive loads result in a decrease in the amplitude and an increase in the latency of several ERP components. A typical example of longer-latency ERP components is the P3 (or P300), an ERP elicited in the process of decision-making. The P3 component is an endogenous ERP component related to cognitive processes such as attention, rather than physical properties of the stimulus that caused it in the first place. It can be divided into two sub-components, the P3a and P3b, which peak at different sites on the scalp. The delay in P3 (sub-)components is around 250--500~ms and depends on the task \citep{picton_1992}. 

\begin{figure}[!ht]
   \centering
   \includegraphics[scale=0.6]{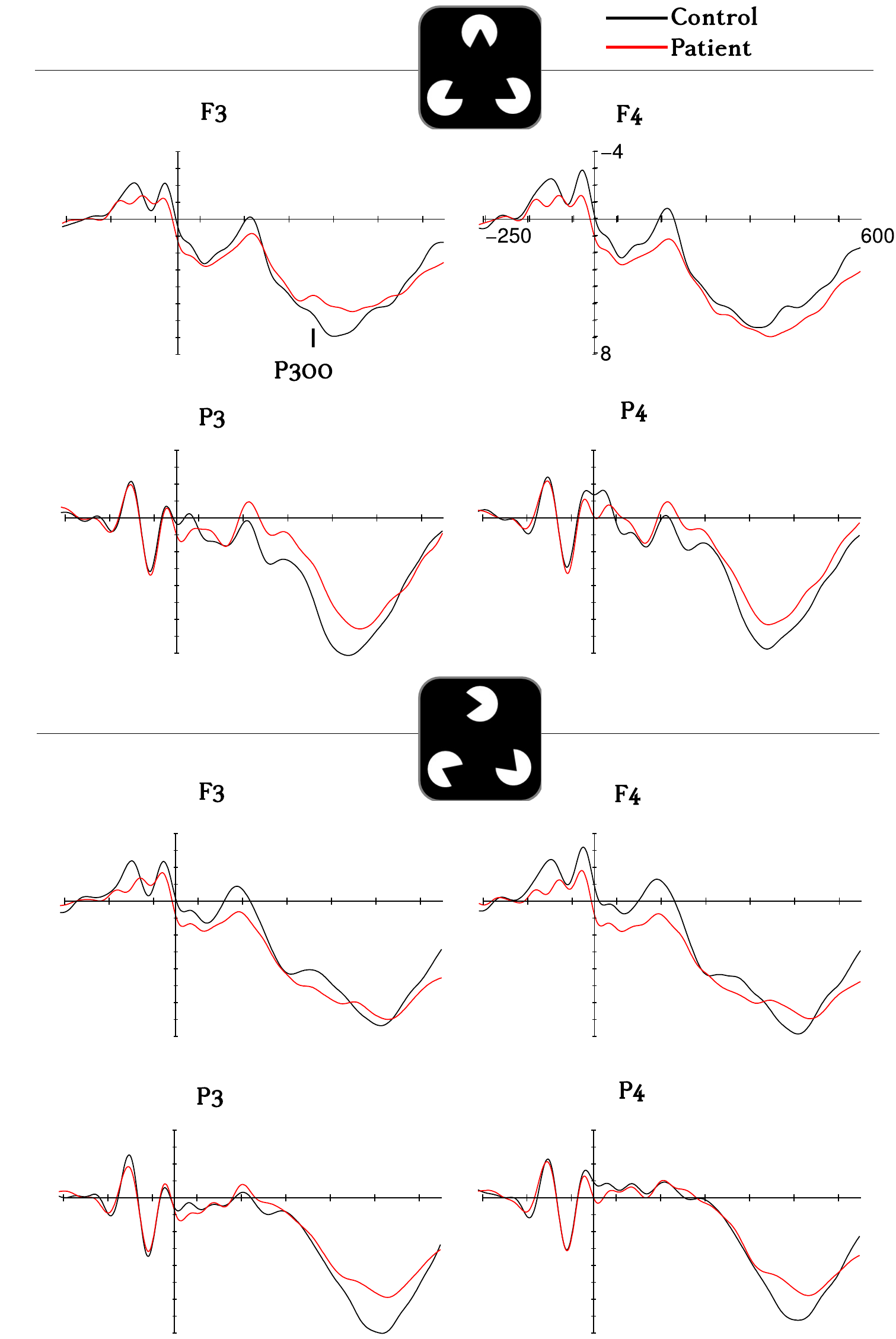}
   \caption{An example of ERPs, from four electrode sites -- F3, F4, P3, and P4 (`F' for frontal, `P' for parietal, `3' for left, and `4' for right) -- in two conditions for a computerised response task (the condition targets are shown in black squares). The P300 is marked in the top-left panel; abscissa (ms) and ordinate ($\mu$V) scales are marked at the top right.}
   \label{fig.ERP}
\end{figure}

Figure \ref{fig.ERP}, taken from reporting on an experiment that contrasted two groups (ADHD patients and healthy controls) across two conditions (illusory contour and no contour) \citep{Cowley2013poster}, illustrates some simple properties of ERPs. Firstly, N100 and P300 waves are clearly visible in all panels. Secondly, the P300 at posterior sites is much less condition-responsive than that at frontal sites. This is logical, given the back-to-front propagation of visual processing in the cortex. Thirdly, the conditions discriminate between the groups: when compared to the controls, ADHD patients show a significantly diminished P300 at F3 (above the left dorsolateral pre-frontal cortex).

Several ERP peaks are visible in responses to sound stimuli recorded while the subject is fully engaged in a primary task with a computer and not paying any attention to the sounds. The auditory P3 brain response is particularly valuable in this connection, since it has been linked to the allocation of attention to surprising, unexpected sounds separate from the primary task. The strength of this response is related to the extent of cognitive resources available at a given moment \citep{Kramer1983}. The auditory P3 is also proposed to index the momentary processing capacity of the relevant individual \citep{Kok2001}.

Somatosensory ERP responses have been proposed as possible indices of cognitive functions, primarily attention, during a task \citep{chen_bi-directional_2014}. Similarly to sounds, somatosensory stimuli can be delivered unexpectedly, without disrupting the primary task excessively. 

ERPs related to a user's own actions have been proposed as markers of user cognitive state. The contingent negative variation (CNV) \citep{Boehm2014} is a preparatory motor potential that is observed prior to movement. Its duration and strength are posited to be related to the planning of the imminent movement. Specifically, inhibition of the response is proportional to the strength of the CNV response even prior to the motor movement after decision-making \citep{Boehm2014}.

Accordingly, the `event' for an ERP may include \textit{responses} to stimuli; indeed, stimulus-event- and response-locked potentials are often analysed in pairs \citep{OConnell2012}. In a study by \citet{OConnell2012}, the response was a subjective decision with respect to continuous (audio and visual) stimuli crossing the perception threshold. This illustrates that ERPs can be derived not only from discrete trials but also from continuous multimodal stimuli.

A further and more field-ready demonstration of ERPs from continuous natural stimuli has been provided by \citet{Poikonen2016}, who extracted musical features from natural sounds and the tango nuevo, techno, and lullaby musical genres. They found that ERPs were detectable in periods of rapid feature increase, especially when such feature `peaks' were preceded by an extended duration of low-feature values. This demonstrates the utility ERPs can have for the practitioner when combined with automated methods of extracting information from natural stimuli.

As described in \citet[sect.3.5]{Cowley2016primer}, cognitive processes are also associated with EEG oscillations, and the relationship between ERPs and oscillations has been investigated by, for example, \citet{Watanabe2002}, who found that event-related gamma-band oscillation and the P3 may share several components of a common larger process related to recognising rare stimuli.
Indeed, if we consider the time course of an ERP as a summation of fixed-frequency waves -- e.g., the P3 is a wave with a 300~ms wavelength -- considering ERPs to represent a brief-time-window snapshot of convolved oscillations is intuitive.

\section{Methods}

The recording for ERP signals is similar to that used for EEG applications, but it additionally requires a temporally very accurate time stamp for the events in question. This implies that the experimenter must account for transmission latencies between the EEG amplifier and the source of the stimulation event. In consequence of the very high precision required, one should also correct for time stamp `jitter' (the slight non-periodicity inherent in any electronic signal) when events are in a stream. 
On the assumption that these preconditions are met, in a typical basic research setting, epochs are combined across hundreds of experiment trials and dozens of participants, to form a \textit{grand average} waveform \citep{Luck2014}.

The process of averaging is a key factor, because the  waveforms characteristic of the ERP may not be seen clearly in any of the individual trials unless they are of large amplitude. This \textit{noisiness} of the individual signals can arise from source mixing and volume conduction, in addition to external contributions such as artefacts. Therefore, the ERP can be considered an idealised representation of the underlying process.

In an applied setting, one should use EEG amplifiers that are lightweight and portable to improve usefulness and affordability; even some consumer-grade devices can be used.
In recent years, there has been a great increase in the commercial availability of ambulatory EEG sensors, partly due to technological advances and also on account of the popularity of the `quantified self' movement.
The sensors available range from lower-cost consumer-oriented models such as the Muse headband, from Interaxon (Toronto, Canada), with fixed electrodes, to expensive medical-grade devices such as the Embla Titanium, from Natus Medical Incorporated (Pleasanton, CA, USA). The Enobio (Neuroelectrics, Barcelona, Spain) amplifier is an intermediate device that is suitable for research purposes and supported in this respect.

Consumer-oriented devices are usually accompanied by software aimed at self-quantification and cognitive enhancement; for instance, the MUSE band comes with an application that teaches meditation techniques. Devices such as this nevertheless usually employ a communication protocol that allows other software to access the device's raw data.
Adding time stamps in the form of triggers, however, may prove challenging, since the solutions currently available for triggering portable EEG recorders do not typically demonstrate ERP-grade temporal accuracy. This necessitates manual verification of the trigger latency and jitter by the experimenter, a time-consuming and technically demanding task. 
In the field of BCI, issues of raw data access or temporal registration have usually been resolved in isolation within labs; however, this is problematic, as BCI research is largely concerned with clinical studies \citep{bamdad_application_2015,ahn_review_2014}.
For these two reasons, most of the relevant reports to date have been from laboratory-based experiments. That said, the impediments to implementation outside the lab seem to be mainly an engineering problem, in terms of both data recording and design of the stimulation events.

\section{Applications}

ERP applications require that a stimulus event of sufficient information-processing novelty evoke a clear brain response. The stimulus may be environmental or be triggered by predefined subject behaviour during an HCI task. This implies one of the following: \textbf{a}) the person is going to be slightly or moderately distracted on occasion by an external stimulus (e.g., sound, a somatosensory stimulus, or light); \textbf{b}) the event is extracted from the task-related behaviour of the person in question \citep{OConnell2012}; \textbf{c}) the event is isolated from a continuous stimulation stream, such as a piece of music \citep{Poikonen2016}.

It should be noted with respect to external events that a secondary task related to external stimuli may annoy, distract, or fatigue the participant \citep{Kramer1991,Trejo1995,Ullsperger2001}. Therefore, external events should be rare enough with respect to the primary task. In addition, the stimuli should be subtle enough not to disturb but salient enough for evoking measurable ERPs. In some cases, the strength of the stimulus, e.g. loudness of a sound, may need to be adjusted for each individual.

Studies showing the effects of cognitive load and processing capacity on the auditory P3 response have been conducted in the field (i.e. in ecologically valid settings, outside the lab) or in flight simulators \citep{Fowler1994,Sirevaag1993}, during simulated driving \citep{Bayliss2000}, and in safety-critical monitoring \citep{Trejo1995,Ullsperger2001}. Furthermore, the P3 has been reliably extracted with freely moving subjects performing an auditory oddball task outside \citep{DeVos2014}, indicating that the relevant paradigm is nearly field-ready.

ERP paradigms can be developed to use stimuli that are elicited by the participants. For instance, the P3 difference between unexpected vs. expected stimuli can be locked to the participant's mouse clicks \citep{Nittono2003}.
More recently, blink-related activity and N2 waves have been found to be predictive of user condition in a study of a simulated logistics-work environment, where the conditions involved physical effort, cognitive effort, and rest \citep{Wascher2014}.

\section{Conclusion}

Recording EEG data with portable devices and in real-world settings is non-invasive, is affordable, and exhibits a high success rate. However, the practical application of ERPs requires suitable hardware/software, which is not currently straightforward to obtain, and demands a clever stimulation protocol that is sufficient to evoke detectable responses yet not interrupt the user.
Once such solutions have been identified, ERPs should constitute a valid approach for workplace HCI. Their use as a control input is still limited by the very low information throughput \citep{Allison2007}, but they possess a clear strength in their long history as a tool used to detect and probe cognitive states. 
We can conclude that using ERPs is to be recommended whenever the application clearly supports the rather strict requirements imposed by this approach.

\bibliographystyle{plainnat}
\bibliography{ch6_erp_bib}

\end{document}